\documentclass[twocolumn,english,amsmath,amssymb,superscriptaddress,nofootinbib]{revtex4-1}
\usepackage[T1]{fontenc}
\usepackage[latin9]{inputenc}
\usepackage{amsmath}
\usepackage{amssymb}
\usepackage{graphicx}
\usepackage{overpic}

\makeatletter
\usepackage[normalem]{ulem}

\usepackage{color}

\makeatother

\usepackage{babel}
\begin{document}
\title{Picocavity-controlled Sub-nanometer Resolved Single Molecule Non-linear Fluorescence}
\author{Siyuan Lyu}
\address{Department of Physics, University of Science and Technology Beijing, Beijing 100083, China}
\author{Yuan Zhang }
\email{yzhuaudipc@zzu.edu.cn}
\address{School of Physics and Microelectronics, Zhengzhou University, Daxue Road 75, Zhengzhou 450052 China}
\author{Yao Zhang}
\address{Hefei National Laboratory for Physical Sciences at the Microscale
and Synergetic Innovation Centre of Quantum Information and Quantum
Physics, University of Science and Technology of China, Hefei, Anhui
230026, China} 
\author{Kainan Chang}
\address{ GPL, State Key Laboratory of Applied Optics, Changchun Institute of Optics Fine Mechanics and Physics, Chinese Academy of Sciences,
Changchun 130033, China}
\author{Guangchao Zheng}
\address{School of Physics and Microelectronics, Zhengzhou University, Daxue Road 75, Zhengzhou 450052 China}
\author{Luxia Wang}
\email{luxiawang@sas.ustb.edu.cn}
\address{Department of Physics, University of Science and Technology Beijing, Beijing 100083, China}

\begin{abstract}
In this article, we address fluorescence of single molecule inside a plasmonic picocavity by proposing a semi-classical theory via combining the macroscopic quantum electrodynamics theory and the open quantum system theory. To gain insights into the experimental results [Nat. Photonics, 14, 693 (2020)], we have further equipped this theory with the classical electromagnetic simulation of the pico-cavity, formed by single atom decorated silver STM tip and a silver substrate,  and the time-dependent density functional theory calculation of zinc phthalocyanine molecule. Our simulations  not only reproduce  the fluorescence spectrum as measured in the experiment, confirming the influence of extreme field confinement afforded by the picocavity, but also reveal Rabi oscillation dynamics and Mollow triplets spectrum for moderate laser excitation. Thus, our study highlights the possibility of coherently manipulating the molecular state and exploring non-linear optical phenomena with the plasmonic picocavity. 
\end{abstract}
\maketitle

\section{Introduction}

Recently proposed concept "picocavity" refers to atomistic
protrusions inside metallic nanocavities \citep{FBenz,CCarnegie,HHShin,QZhou,JLee,TallaridaN,JLee2019,BDoppagne,BYang,ARoslawska,ZHe,MUrbieta} formed by metallic nanoparticle on-film constructs \citep{FBenz,CCarnegie,HHShin,QZhou}, STM tip on-film structures \citep{JLee,TallaridaN,JLee2019,BDoppagne,BYang,ARoslawska,ZHe} or metallic nanoparticle dimers \citep{MUrbieta}. The atomic protrusions enhance the local field in an atomistic scale due to a non-resonant lighting-rod effect \citep{MUrbieta} over hundreds fold of enhanced local field afforded by the resonant gap plasmon of the metallic nanocavities. 
Since the extreme confined field can be smaller than the spatial extension of molecules, its interaction with the molecules can not be addressed with the typical form of light-matter
interaction, which assumes large extension of the electromagnetic
field over the molecule, i.e. the dipole approximation. This particular situation requires us to go beyond and address the influence of the atomistic local field on the light-matter interaction \citep{TNeuman,YZhang}, and on the resulting optical phenomena,
such as surface-enhanced Raman scattering \citep{FBenz,JLee,CCarnegie,HHShin} and surface-enhanced fluorescence \citep{BDoppagne,BYang,ZHe}. 

\begin{figure}
\begin{centering}
\includegraphics[scale=0.9]{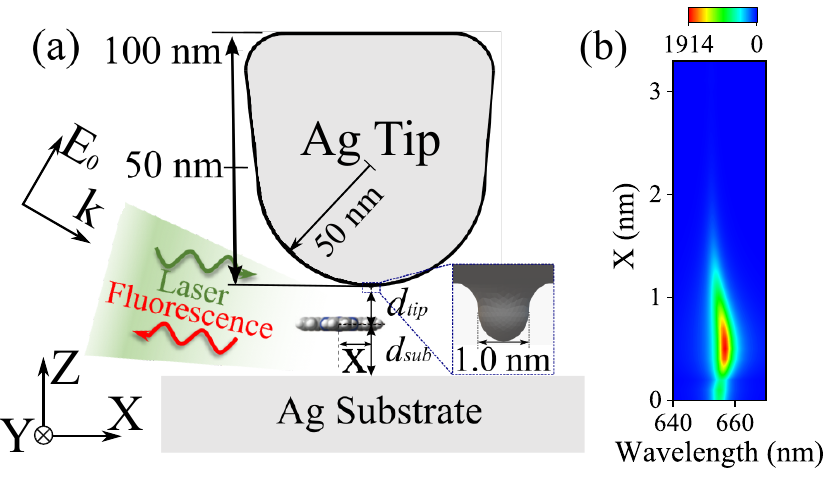}
\par\end{centering}
\caption{\label{fig:1} Fluorescence from single molecule in a picocavity. (a) shows the STM-based picocavity, where the STM tip is modeled as a cone over a hemisphere with size marked and the atomic protrusion is modeled as a sphere of 1 nm diameter (inset), and single zinc phthalocyanine molecule laying horizontally, whose center departs vertically from the substrate by a distance $d_{sub}=1.4$ nm (enabled by a NCl spacer in the experiment), and from the STM apex by $d_{tip}$, and is horizontally away from the apex by X. The picocavity is illuminated by a plane-wave field with amplitude $\bf{E}_0$ and wavevector $\bf{k}$, which is thirty degrees from the substrate normal, and the fluorescence signal is determined in the reversed path. (b) shows the calculated fluorescence spectrum according to our theory (for $d_{tip} =0.4$ nm).   } 
\end{figure}

In Ref. \citep{BYang}, B. Yang et al, reported experimentally  fluorescence imaging with subnanometer resolution from single zinc phthalocyanine (ZnPc) molecule inside a STM-based metallic picocavity (Fig. \ref{fig:1}a). To verify the involvement of the picocavity, the authors have also studied the influence of the STM tip-molecule distance on the surface-enhanced fluorescence signal, the fluorescence linewidth and line shifts. However, in the corresponding theoretical study, the different quantities are studied with separated formulas, and the internal connection between them is lost.

In the present article, we go beyond the theory developed in  \citep{BYang}, and propose a semi-classical theory by combining the macroscopic quantum electrodynamics theory \citep{NRivera,Scheel} and the open quantum system theory \citep{HPBreuer}. Our theory units different quantities in a coherent manner, and more importantly allows us to calculate the fluorescence spectrum, as measured directly in the experiment, which thus provide more insights. Furthermore, we equip our theory with the classical electromagnetic simulations \citep{UHohenester,JWaxenegger} and the time-dependent density functional theory (TDDFT) calculation to investigate the response of the ZnPc molecule to the laser excitation with varying wavelength and intensity. Our simulations not only reproduce the fluorescence spectra measured in the experiment, see Fig. \ref{fig:1}b for an example, but also predict the possibility of observing the coherent Rabi-oscillation in the excited state population dynamics and the Mollow side-peaks spectrum under moderate laser excitation. 

Our article is organized as follows. In the following section, we present our semi-classical theory, including the quantum master equation for the system dynamics and the formula to compute the fluorescence spectra. In Sec. \ref{sec:plasmon} and \ref{sec:TDDFT}, we study the plasmonic response of the picocavity via classical electromagnetic simulations, and the excitation energy and transition current density of the ZnPc molecule via the TDDFT calculation. In Sec. \ref{sec:couplings}, we investigate the molecule-local field coupling, the plasmonic Lamb shift and the Purcell-enhanced decay rate, which are then utilized in Sec. \ref{sec:spectrum} to investigate the influence of tip-molecule distance, laser wavelength and laser intensity on the fluorescence spectrum. In the end, we conclude our work and comment on the possible extensions in future.

\section{Quantum Master Equation and Fluorescence Spectrum Formula \label{sec:sct}}

In the Appendix \ref{sec:molpic}, we have considered the interaction
between single molecule and quantized electromagnetic field of the
picocavity according to macroscopic quantum electrodynamics theory \citep{NRivera,Scheel}, and derived a quantum master equation for the molecule by tracing out the electromagnetic field reservoir. Including the excitation of the molecule and other
dissipation processes, we arrive at the following quantum master equation
\begin{align}
\frac{\partial}{\partial t}\hat{\rho} & =-\frac{i}{\hbar}\left[\hat{H}_{ele}+\hat{H}_{las},\hat{\rho}\right]\nonumber \\
& +\frac{1}{2}\left(\Gamma+\gamma\right)\left(\left[\hat{\sigma}^{-}\hat{\rho},\hat{\sigma}^{\dagger}\right]+\left[\hat{\sigma}^{-},\hat{\rho}\hat{\sigma}^{\dagger}\right]\right) \nonumber \\
& +\frac{1}{4}\chi\left(\left[\hat{\sigma}^{z},\hat{\rho}\hat{\sigma}^{z}\right]+\left[\hat{\sigma}^{z}\hat{\rho},\hat{\sigma}^{z}\right]\right).
\label{eq:meq}
\end{align}
We consider the electronic ground and excited states of the
molecule, and treat them as two-level system via the Hamiltonian
$H_{ele}=\hbar\left[\left(\omega_{eg}-\Omega/2\right)/2\right]\hat{\sigma}^{z},$
where $\omega_{eg}-\Omega/2$ is the transition frequency (accounting
for the plasmonic Lamb shift $\Omega/2$), and $\hat{\sigma}^{z}$ is the Pauli operator. We treat the optical excitation of the molecule in a semi-classical way, and introduce the driving Hamiltonian $H_{las}=\hbar\left(ve^{-i\omega_{l}t}\hat{\sigma}^{\dagger}+v^{*}e^{i\omega_{l}t}\hat{\sigma}^{-}\right)$,
where $\hat{\sigma}^{\dagger}$ and $\hat{\sigma}^{-}$ are the raising and lowering operators, respectively. The coupling coefficient $\hbar v=-\frac{ie}{\omega_{eg}}\int d^3\mathbf{r} \mathbf{j}_{eg}\left(\mathbf{r}\right)\cdot\mathbf{E}_{loc}\left(\mathbf{r},\omega_{l}\right)$
is determined by the local electric field $\mathbf{E}_{loc}\left(\mathbf{r},\omega_{l}\right)$
at the position $\mathbf{r}$ excited by a laser with frequency $\omega_{l}$.
Here, $e$ is the elementary charge. The transition current density
is defined as
\begin{equation}
\mathbf{j}_{eg}\left(\mathbf{r}\right)=-\frac{i\hbar}{2m_{e}}\left[\Psi_{e}^{*}\left(\mathbf{r}\right)\nabla\Psi_{g}\left(\mathbf{r}\right)-\Psi_{g}^{*}\left(\mathbf{r}\right)\nabla\Psi_{e}\left(\mathbf{r}\right)\right] \label{eq:tran_curr}
\end{equation}
with the electron mass $m_{e}$, the wavefunctions $\Psi_{g}\left(\mathbf{r}\right),\Psi_{e}\left(\mathbf{r}\right)$
of the electronic ground and excited state, respectively. The both wavefunctions can be computed through the TDDFT calculation, see Sec. \ref{sec:TDDFT}.

The remaining terms in Eq. (\ref{eq:meq}) describe the dissipation of the molecule, where the second line describes the decay from the excited state with the total rate $\Gamma+\gamma$, where the rate $\Gamma$ is due to the coupling with the picocavity, known as the Purcell-enhanced decay rate, and the third line describes the dephasing rate $\chi$ of the molecular transition. The transition frequency shift $\Omega/2$ and the Purcell-enhanced decay rate $\Gamma$ can be computed with 
\begin{align}
\Omega & =\frac{2e^{2}}{\hbar\epsilon_{0}c^{2}}\int d^3\mathbf{r}\int d^3\mathbf{r}'\mathbf{j}_{eg}\left(\mathbf{r}\right)\cdot\mathrm{Re}\overleftrightarrow{G}\left(\mathbf{r},\mathbf{r}';\omega_{eg}\right)\cdot\mathbf{j}_{eg}^{*}\left(\mathbf{r}'\right),\label{eq:Omega}\\
\Gamma & =\frac{2e^{2}}{\hbar\epsilon_{0}c^{2}}\int d^3\mathbf{r}\int d^3\mathbf{r}'\mathbf{j}_{eg}\left(\mathbf{r}\right)\cdot\mathrm{Im}\overleftrightarrow{G}\left(\mathbf{r},\mathbf{r}';\omega_{eg}\right)\cdot\mathbf{j}_{eg}^{*}\left(\mathbf{r}'\right).\label{eq:Gamma}
\end{align}
Here, $\epsilon_{0},c$ are the permittivity and the light speed in vacuum, respectively, and $\overleftrightarrow{G}\left(\mathbf{r},\mathbf{r}';\omega\right)$ is the classical dyadic Green's function. 
The function $\overleftrightarrow{G}\left(\mathbf{r},\mathbf{r}';\omega\right)$
and $\mathbf{E}_{loc}\left(\mathbf{r},\omega_{l}\right)$ can be computed
through classical electromagnetic simulations, see Sec. \ref{sec:plasmon}.

\begin{figure*}
\begin{centering}
\includegraphics[scale=0.9]{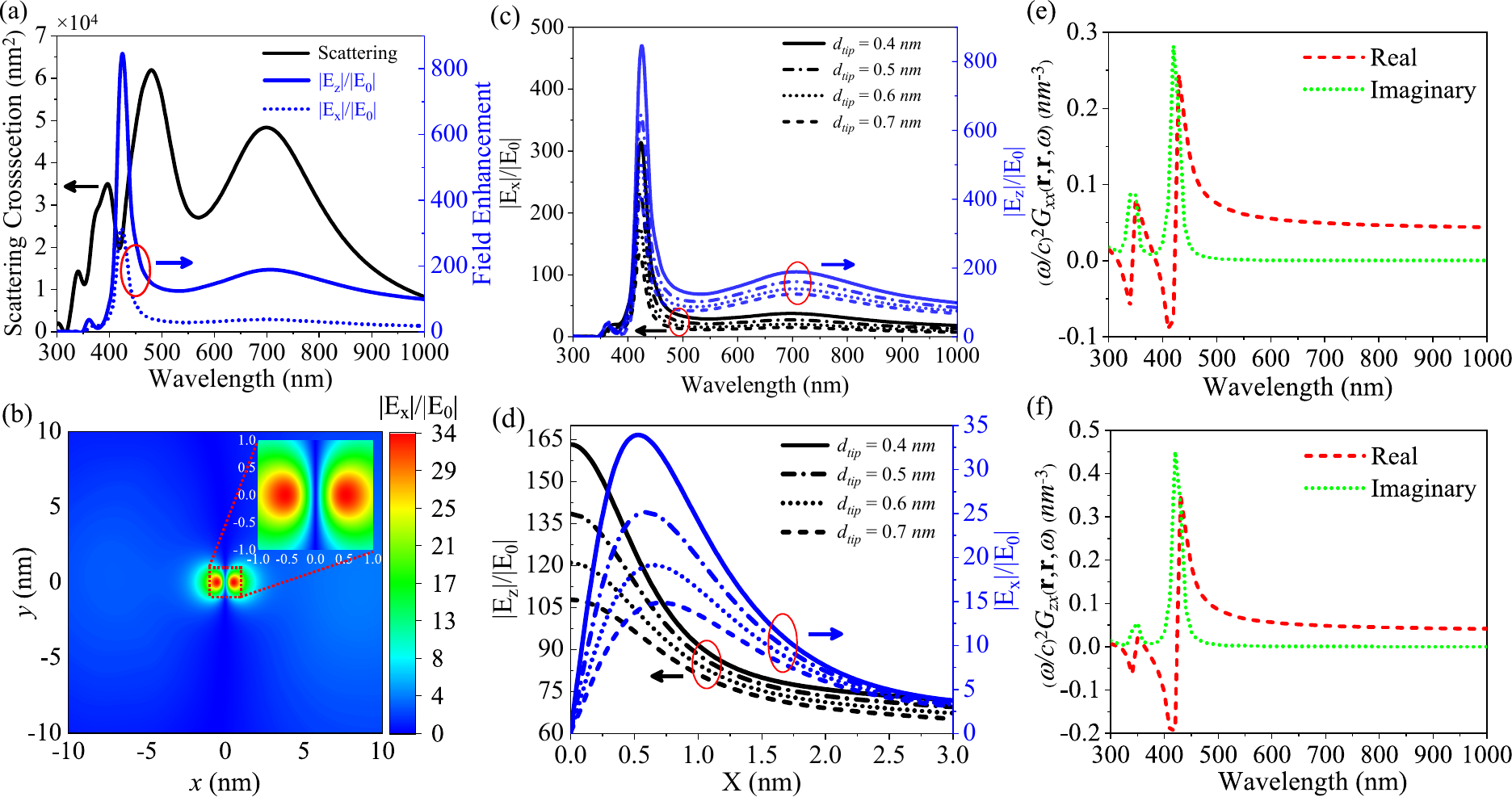}
\par\end{centering}
\caption{ \label{fig:2} Plasmonic response of the picocavity. (a) shows the far-field scattering cross-section (black solid line, left axis), and the enhancement of local field z- and x-component (blue solid and dashed line, right axis) as a function of wavelength of plane-wave illumination.  Here, the z- and x-component are evaluated at the molecular center and a point about $0.5$ nm away from the molecular center, respectively. (b) shows the map of the x-component field enhancement in the middle of the picocavity for the $633$ nm plane-wave illumination. (c) shows the similar results as the near-field component in (a) except that the STM tip-molecule distance  $d_{\rm tip}$ varies from $0.4$ nm to $0.7$ nm. (d) shows the field enhancement of the z-component (black lines, left axis) and x-component (blue lines, right axis) as a function of the tip-molecule horizontal distance for increasing  $d_{\rm tip}$. (e) and (f) show the real part (red dashed lines) and the imaginary part (green dotted lines) of the scattered dyadic Green's function for the STM tip about $0.5$ nm away from the molecular center.} 
\end{figure*}

The differential power $\frac{dW}{d\Omega}\left(\omega\right)$ at the position $\mathbf{r}_{d}$ of a detector can be computed with
the Fourier transformation of the correlation function of the electric field operators $\hat{\mathbf{E}}\left(\mathbf{r}_{d},\tau\right)$. Relating this electric field operator with the lowering operator
$\hat{\sigma}^{-}\left(\tau\right)$ of the molecule through the dyadic Green's function in the Markov approximation, we obtain finally the expression $\frac{dW}{d\Omega}\left(\omega\right)\approx K\mathrm{Re}\int_{0}^{\infty}d\tau e^{i\omega\tau}\mathrm{tr}\left\{ \hat{\sigma}^{-}\hat{\varrho}\left(\tau\right)\right\} ,$
and $\hat{\varrho}\left(\tau\right)$ satisfies the same equation (\ref{eq:meq}) as $\hat{\rho}$ with however the initial condition $\hat{\varrho}\left(\tau\right)=\hat{\rho}_{ss}\hat{\sigma}^{\dagger}$, where $\hat{\rho}_{ss}$ is the density operator at steady-state.
In this expression, the propagation factor is defined as
\begin{align}
K & =\frac{r^{2}\omega_{eg}^{2}e^{2}}{4\pi^{2}\epsilon_{0}c^{3}}\int d^3\mathbf{r}''\int d^3\mathbf{r}'\left[\overleftrightarrow{G}^{*}\left(\mathbf{r}_{d},\mathbf{r}'';\omega_{eg}\right)\cdot\mathbf{j}_{eg}\left(\mathbf{r}''\right)\right]\nonumber \\
 & \cdot\left[\overleftrightarrow{G}\left(\mathbf{r}_{d},\mathbf{r}';\omega_{eg}\right)\cdot\mathbf{j}_{eg}^{*}\left(\mathbf{r}'\right)\right],\label{eq:Propagation}
\end{align}
where $r$ is the distance between the molecular center and the detector. 

As a comparison, we consider also the corresponding expressions under the dipole approximation.
It assumes that the spatial variation of the quantities $\mathbf{E}_{loc}\left(\mathbf{r},\omega_{l}\right)$
, $\overleftrightarrow{G}\left(\mathbf{r},\mathbf{r}';\omega\right)$
and $\overleftrightarrow{G}\left(\mathbf{r}_{d},\mathbf{r}';\omega_{eg}\right)$, is much larger than the spatial extension of the molecular transition density
$\mathbf{j}_{eg}\left(\mathbf{r}\right)$. As a result, in Eqs. (\ref{eq:Omega}-\ref{eq:Propagation}), we can replace the
position argument $\mathbf{r},\mathbf{r}',\mathbf{r}''$  by the molecular center position $\mathbf{r}_{m}$, the integration $\frac{ie}{\omega_{eg}}\int d^3\mathbf{r}\mathbf{j}_{eg}\left(\mathbf{r}\right)$ by  the transition dipole moment $\mathbf{d}_{eg}$, and the dyadic Green's functions by  $\overleftrightarrow{G}\left(\mathbf{r}_{m},\mathbf{r}_{m};\omega_{eg}\right), \overleftrightarrow{G}\left(\mathbf{r}_{d},\mathbf{r}_{m};\omega_{eg}\right)$.

\section{Plasmonic Response of STM-based Picocavity \label{sec:plasmon}}

To study the plasmonic response of the STM-based picocavity, as shown in Fig.\ref{fig:1}a, we carry out the electromagnetic simulations by solving the Maxwell's equation with boundary element method (BEM) \citep{FJGDAbajo,FJGDAbajo1} as implemented in metallic nanoparticle BEM toolkit \citep{UHohenester,JWaxenegger}, and utilize the permittivity of silver given by Johnson-Christy \citep{PBJohnson}. 

To mimic the experiment, we illuminate the plasmonic picocavity with a p-polarized plane-wave field at an incident angle of $30^{\circ}$ with respect to the substrate normal (Fig. \ref{fig:1}a). Fig. \ref{fig:2}a shows the computed far-field scattering cross section (black solid line, left axis), and the near-field enhancement of the z-component at the center of picocavity (blue solid line, right axis), and of the x-component at a point $0.5$ nm away horizontally from the the center (blue dashed line, right axis).  The scattering spectrum shows four peaks at $700$ nm, $480$ nm, $390$ nm and $340$ nm, while the near-field enhancement spectra show mainly two peaks at $707$ nm, $425$ nm. 
The peaks at $700$ nm and $480$ nm can be attributed to the (10) and (20) mode according to the nomenclature proposed in \citep{KongsuwanN2020},  and the peaks are due to the higher order plasmonic modes.  The maximum of the peaks is about $850$, $190$ for the field z-component, and is however about three times smaller for the field x-component. 

 As a comparison, we have also calculated the spectra for a corresponding nano-cavity (without the atomic protrusion), see Fig. \ref{fig:2a_appendix} in Appendix \ref{sec:snr}, and found that the scattering spectrum is the same but the near-field enhancement is much smaller, and also does not show the sharp peak at around $410$ nm. This comparison indicates that the atomic protrusion affects mainly the local field but not the far-field \citep{FBenz}, and the sharp peak at around $410$ nm might be caused by the Fabry-Perot-like mode formed in the atomic protrusion \cite{TWu}. 
 
To further understand the plasmonic response, we have also computed the near-field map in the middle plane of the picocavity for the $633$ nm plane-wave illumination, see Fig. \ref{fig:2}b. The x-component of the local field is minimal at the center, and maximal at two points about $0.6$ nm away from the center along the x-axis. The y-component shows similar pattern except that the maximum occurs along the y-axis (Fig. \ref{fig:2b_appendix}a in Appendix \ref{sec:snr}). The both components concentrate in an area of $5$ nm size. The field z-component shows the maximum at the center, and is much larger compared to other field components, as well as concentrate in an area of $1$ nm size over a broad background (Fig. \ref{fig:2b_appendix}b in Appendix \ref{sec:snr}).   

Since we shall study the change of fluorescence as the STM tip moves away from the molecular center later on, we examine here the influence of the tip-molecule distance $d_{\rm tip}$ on the local field enhancement. Fig. \ref{fig:2}c shows that the field enhancement reduces by about 2 times as  $d_{\rm tip}$ increases from $0.4$ nm to $0.7$  nm, and the wavelength of the plasmonic resonance blue-shifts slightly.  Furthermore,  Fig. \ref{fig:2}d shows that accompanying with the reduced enhancement, the spatial extension of the field component becomes also slightly broadened. Here, we focus on the field x-component, since it will affect dominantly  the fluorescence, as shown later on.

After examining the near-field enhancement,  we study now the response of the dyadic Green's function of the picocavity. Fig. \ref{fig:2}e,f show the real part (red dashed line) and the imaginary part (green dotted line) of the xx- and zx-component of the dyadic Green's tensor. The yx-component of the dyadic Green's tensor is very small (Fig. \ref{fig:2b_appendix}c in Appendix \ref{sec:snr}) and thus is not shown here. The imaginary part shows two peaks at the same wavelength of maximal field enhancement but does no show peak around $700$ nm. Here, we see also two Fano-features around $420$ nm and $380$ nm in the real part (red dashed line) of the dyadic Green's function. These results indicate that the dyadic Green's function is mainly determined by the gap structure formed by the atomic protrusion and the substrate \citep{YZhang1}.

\begin{figure}
\begin{centering}
\includegraphics[scale=0.65]{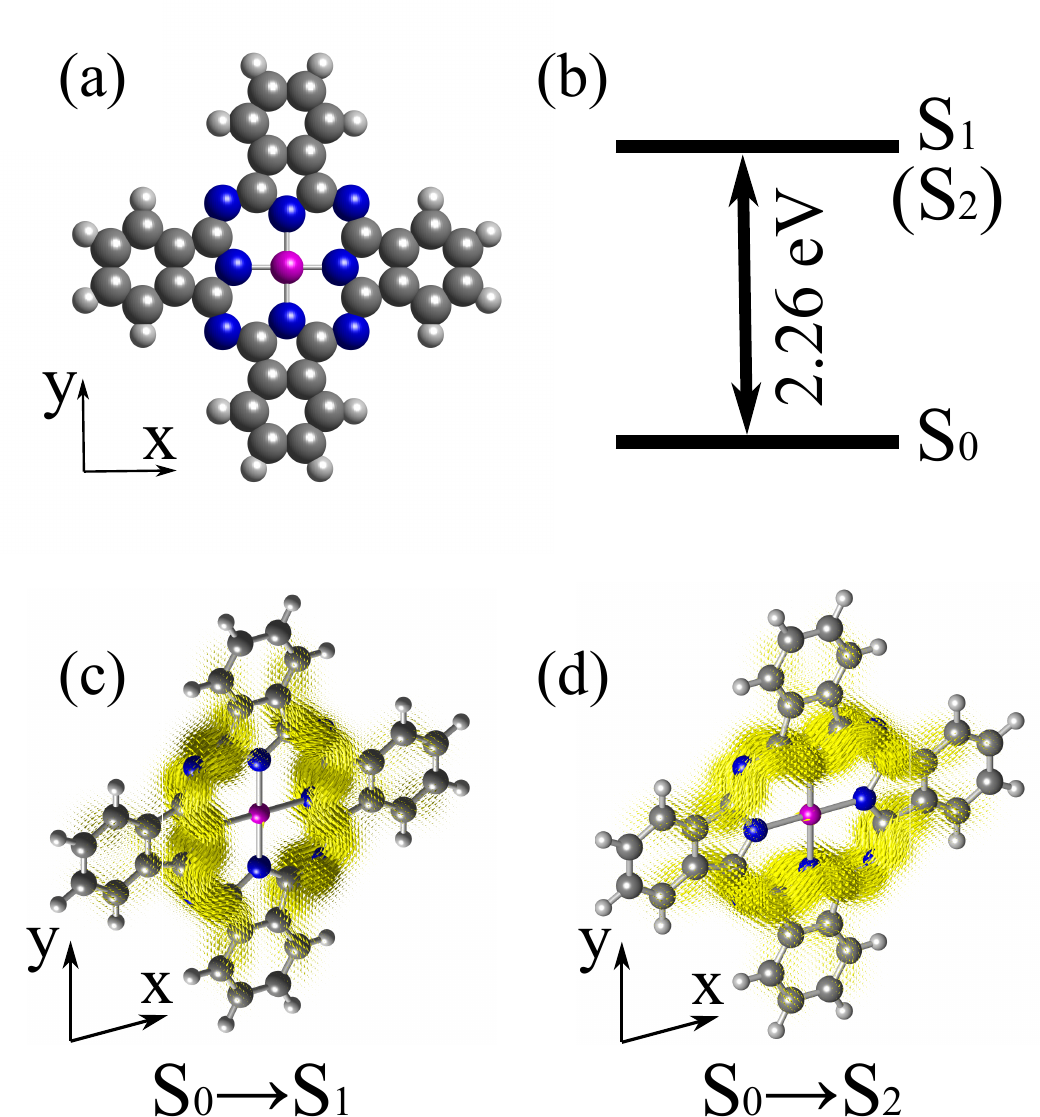}
\par\end{centering}
\caption{ \label{fig:3} TDDFT calculations of the ZnPc molecule. (a) shows the atomic structure of the molecule, where white, gray, blue and pink spheres represent the hydrogen, carbon, nitrogen and zinc atoms, respectively. (b) shows energy level diagram with two degenerate excited states. (c) and (d) show the transition currents of the two excitonic transitions with vectors around the bonds of the molecule. }
\end{figure}

\section{Time-Dependent Density Functional Calculations of ZnPc Molecule \label{sec:TDDFT}}

In this section, we present the TDDFT calculations of the ZnPc molecule (Fig. \ref{fig:3}). To obtain these results, we firstly optimize the structure of the ZnPc molecule (Fig. \ref{fig:3}a), and then calculate the electronic ground and excited states by using Gaussian 09 program \citep{MJFrisch} with B3LYP functional and 6-31G(d) basis set. We identify that the transitions HOMO $\to$  LUMO (LUMO+1) contribute mostly to the ${\rm S}_0 \to {\rm S}_1$ (${\rm S}_2$) excitonic transitions (Fig. \ref{fig:3}b) with the same excitation energy $2.26$ eV, which is  smaller than the value $1.90$ eV  measured in the experiment since the TDDFT often overestimates the transition energy. In the following, we will utilize the measured wavelength in our simulation to ensure the comparison with the experimental results. To calculate the transition current ${\bf j}_{eg}(\bf{r})$ according to Eq. (\ref{eq:tran_curr}), we obtain the wave-functions $\Psi_g(\bf{r})$, $\Psi_e(\bf{r})$  of the HOMO, LUMO (LUMO+1) level, which are normally given by $\Psi_{g(e)}({\bf r}) = \sum_i C_i \chi_i ({\bf r})$ with the Gaussian-type $i$-th basis function and the expansion coefficients $C_i$, and calculate the spatial derivative analytically as $\nabla \Psi_{g(e)}({\bf r}) = \sum_i C_i \nabla \chi_i ({\bf r})$.  Furthermore, using the relation $\mathbf{d}_{eg}=\frac{ie}{\omega_{eg}}\int d^3\mathbf{r}\mathbf{j}_{eg}\left(\mathbf{r}\right)$, we calculate also the transition dipole moment  $\mathbf{d}_{eg}$  and then verify our calculation by comparing the calculated values with those given in the TDDFT output file. 

The calculated wave-functions and the transition current ${\bf j}_{eg}(\bf{r})$ are already presented in our previous article \citep{YZhang}. To gain more insights into ${\bf j}_{eg}(\bf{r})$, we show the transition current as vectors around the nitrogen-carbon bonds of the molecule (Fig. \ref{fig:3}c and d). We find that for the ${\rm S}_0 \to {\rm S}_1$ transition the transition current flows upwards along the y-axis, while for the ${\rm S}_0 \to {\rm S}_2$ transition the transition current flows rightwards along the x-axis. Thus, the transition dipole moment,  obtained by integrating the transition current over the space, points along the y-axis and x-axis for the ${\rm S}_0 \to {\rm S}_1$ and ${\rm S}_0 \to {\rm S}_2$ transition, respectively.

\begin{figure}
\begin{centering}
\includegraphics[scale=0.85]{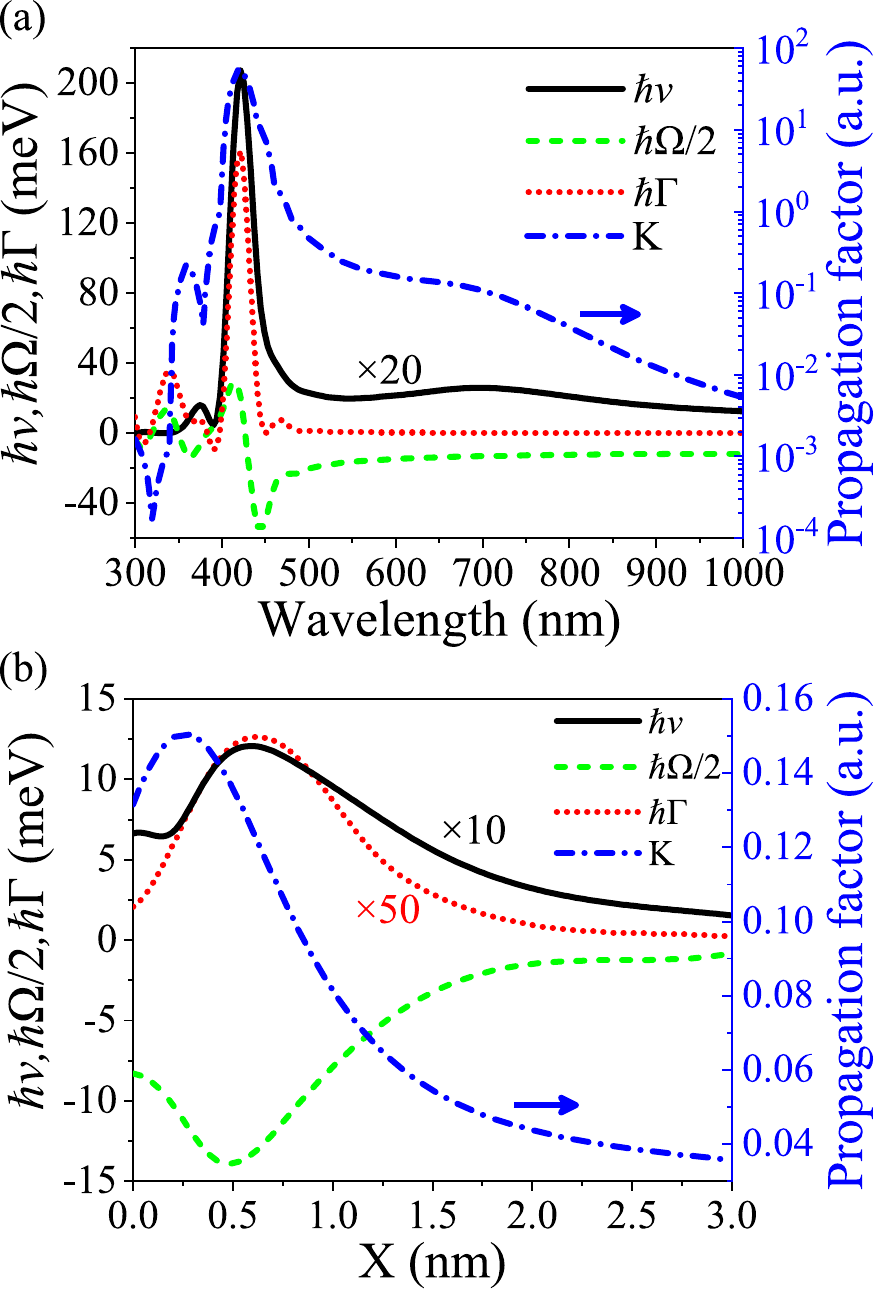}
\end{centering}
\caption{\label{fig:couplings} ZnPc molecule-picocavity couplings for the ${\rm S}_0 \to {\rm S}_2$ excitonic transition. (a) shows the molecule-local field coupling $\hbar v$ (black solid line) for the laser power $10^2 \mu W/\mu m^2$, the plasmonic Lamb shift $\hbar \Omega/2$  (green dashed line), and the Purcell-enhanced decay rate $\hbar\Gamma$  (red dotted line), and the propagation factor $K$ (blue dash-dotted line) as a function of wavelength for the STM tip about $0.5$  nm horizontally away from the molecular center, where the vertical black dashed line shows the wavelength of the molecular transition. (b) shows the change of  $\hbar v$,  $\hbar \Omega/2$,  $\hbar\Gamma$, $K$ as the STM tip moves horizontally away from the molecule along the x-axis for the ZnPc excitonic transition at the wavelength $652$ nm. }
\end{figure}

\section{Molecule-Local Field Coupling, Plasmonic Lamb Shift and Purcell-enhanced Decay Rate \label{sec:couplings}}

We can now combine the quantities calculated in previous sections to determine the molecule-picocavity couplings (Fig. \ref{fig:couplings}), which include the molecule-local field coupling $\hbar v$ for given laser intensity  $I_{\rm las} = 10^2 \mu W/\mu m^2$ (black solid line), the plasmonic Lamb shift $\hbar \Omega/2$ (green dashed line) and the Purcell-enhanced decay rate $\hbar\Gamma$ (red dotted line), and the propagation factor $K$ (blue dash-dotted line). Here, we focus on the ${\rm S}_0 \to {\rm S}_2$ transition since it dominates the fluorescence under the detection condition as considered here (see Fig. \ref{fig:5_appendix} in Appendix \ref{sec:snr}). We find that $\hbar v$ follows the shape of the near-field enhancement, and reaches the maximal value around $10$ meV at the wavelength of $420$ nm. $\hbar \Omega/2$ follows the shape of the real part of the dyadic Green's function, and changes in the range of $[-50,30 { \rm meV}]$. $\hbar \Gamma$ follows the shape of the imaginary part of that function, and varies in the range of $[0,160 {\rm meV}]$. For the wavelength $652$ nm of the ZnPc excitonic transition, we obtain $\hbar v = 1.2$ meV, $\hbar \Omega/2 = -12.5$ meV, $\hbar \Gamma = 0.2$ meV.

Furthermore, we investigate with Fig. \ref{fig:couplings}b how the molecule-local field coupling $\hbar v$, the plasmonic Lamb shift  $\hbar \Omega/2$, the Purcell-enhanced decay rate  $\hbar\Gamma$  and the propagation factor $K$ change with the STM tip-molecule distance X in the horizontal direction. We see that the absolute value of these quantities increase first and then decreases with the increasing distance X. More precisely, the former three quantities reach their maxima around $1.2$ meV, $-14.5$ meV, $0.25$ meV for the distance X around $0.6, 0.5, 0.6$ nm, while  the propagation factor reaches the maximum at the distance X around $0.25$ nm. As a reference, we have also calculated these parameters by modelling the molecule as a point in the dipole approximation (see Fig. \ref{fig:4_appendix}a,b in Appendix \ref{sec:snr}), and found similar results except that the values are relatively larger, and the Lamb shift and the propagation factor reach the maximum at zero distance.

\section{Picocavity-controlled Fluorescence Spectra \label{sec:spectrum}}

We are now in the position to study the picocavity-controlled non-linear fluorescence of single ZnPc molecule. In Subsec. \ref{sec:fsdistance}, we will investigate the change of the fluorescence spectrum as the STM tip moves horizontally and vertically, and demonstrate that our calculations can reproduce the experimental results. In Subsec. \ref{sec:fsintensity}, we go beyond the experiment, and explore the influence of the laser wavelength and laser intensity on the molecular dynamics and the fluorescence spectrum. 

\begin{figure}
\begin{centering}
\includegraphics[scale=0.75]{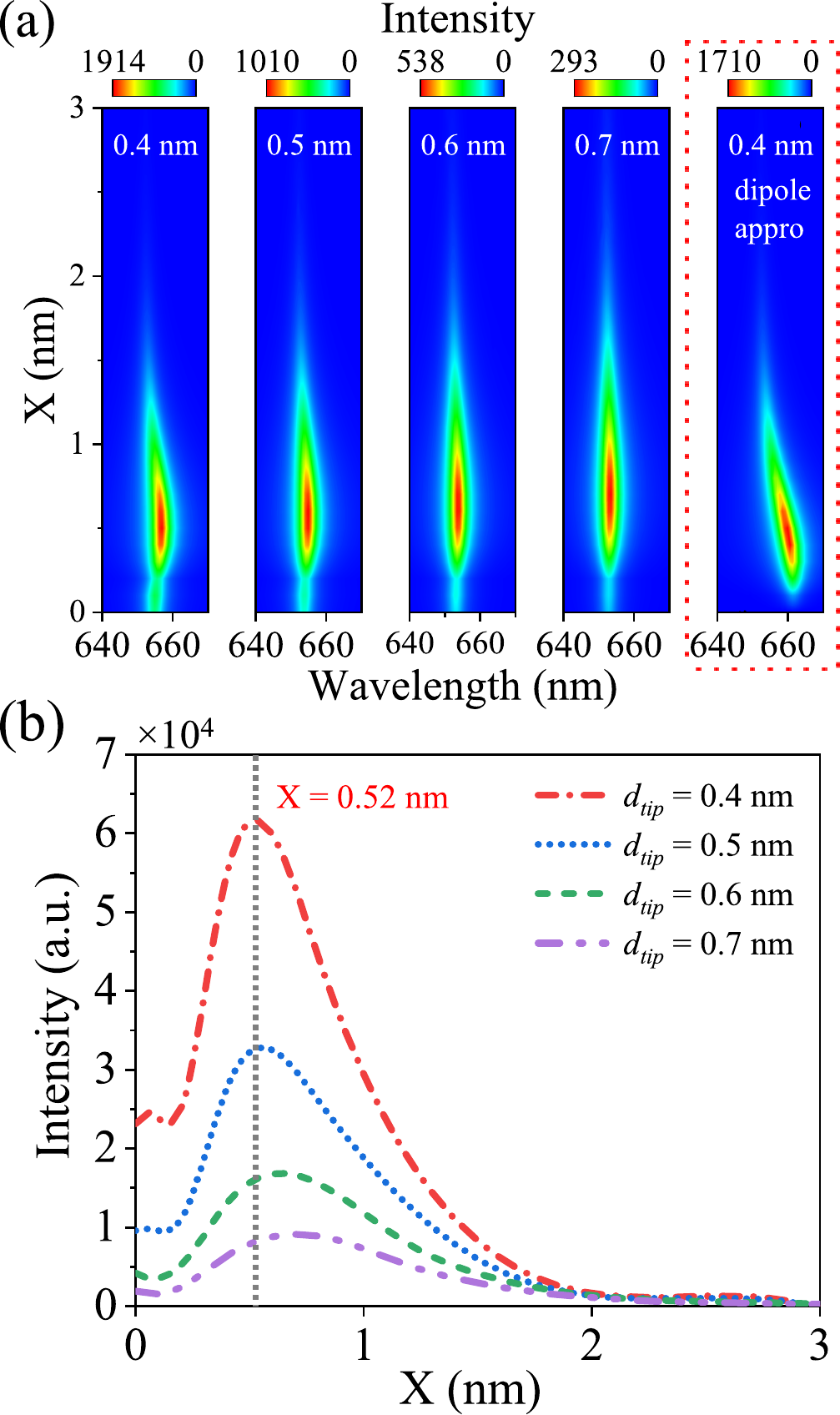}
\par\end{centering}
\caption{\label{fig:5} Picocavity-controlled fluorescence of single ZnPc molecule. (a) shows the evolution of fluorescence spectra as a function of the tip-molecule horizontal distance X for the increased vertical distance $d_{tip}$  from $0.4$ nm to $0.7$ nm (from first to fourth panels), and the result for $d_{tip}=0.4$ nm within the dipole approximation (rightmost panel). (b) shows the integrated fluorescence signal as a function of X for increasing $d_{tip}$  (from top to bottom curves).  In all the simulations, the laser wavelength and intensity are $633$ nm and $I_{las} = 10^2 \mu W/\mu m^2$, respectively, and the dephasing rate and the intrinsic decay rate are assumed as $\hbar \chi = 2.3$ meV and $\hbar \gamma = 8.2 $ meV.} 
\end{figure}

\subsection{Fluorescence Spectra for Different Tip-Molecule Distances \label{sec:fsdistance}}

In the experiment \citep{BYang},  B. Yang et al. had investigated the dependence of the fluorescence spectrum on the tip-molecule distance to verify that the sub-nanometre resolution of fluorescence mapping is afforded by the extremely confined field inside the plasmonic picocavity. Thus, to verify the validity of our theory of the picocavity-controlled fluorescence, in this section, we calculate also the fluorescence spectrum for the different tip-molecule distance for the ${\rm S}_0 \to {\rm S}_2$ excitonic transition (Fig. \ref{fig:5}). We have also computed the fluorescence spectrum for  ${\rm S}_0 \to {\rm S}_1$ transition (Fig. \ref{fig:4_appendix}c in Appendix \ref{sec:snr}) and found that it is orders of magnitude smaller due to the smaller propagation factor for the detection condition as considered here.

Fig. \ref{fig:5}a shows that as the tip-molecule horizontal distance X increases from $0$ to $3$ nm, the fluorescence spectra blue-shift slightly, and their maximum increases firstly and then decays to zero, as well as the distance for the vanishing spectrum increases. These results agree qualitatively with Fig. \ref{fig:2}b in Ref. \citep{BYang}. As a comparison, we show also the result within the dipole approximation (the rightmost panel of Fig. \ref{fig:5}a), and find a larger red-shift of the spectrum for smaller horizontal distances, which is caused by the overestimated plasmonic Lamb shift within the dipole approximation, as explained in the previous section.

To further quantify the influence of the tip-molecule distance, we study the fluorescence intensity $I_{flu}$ integrated over the wavelength range $[648,662]$ nm as a function of the tip-molecule horizontal distance X for different tip-molecule vertical distances $d_{tip}$ (Fig. \ref{fig:5}b). We can see that all the lines increase first and then decrease with the increasing X. In that figure, we see also that the distance X to reach the maximum increases with the increasing  $d_{tip}$. All these results agree qualitatively with Fig. 2b,c in Ref. \citep{BYang}.

\begin{figure*}
\begin{centering}
\includegraphics[scale=0.75]{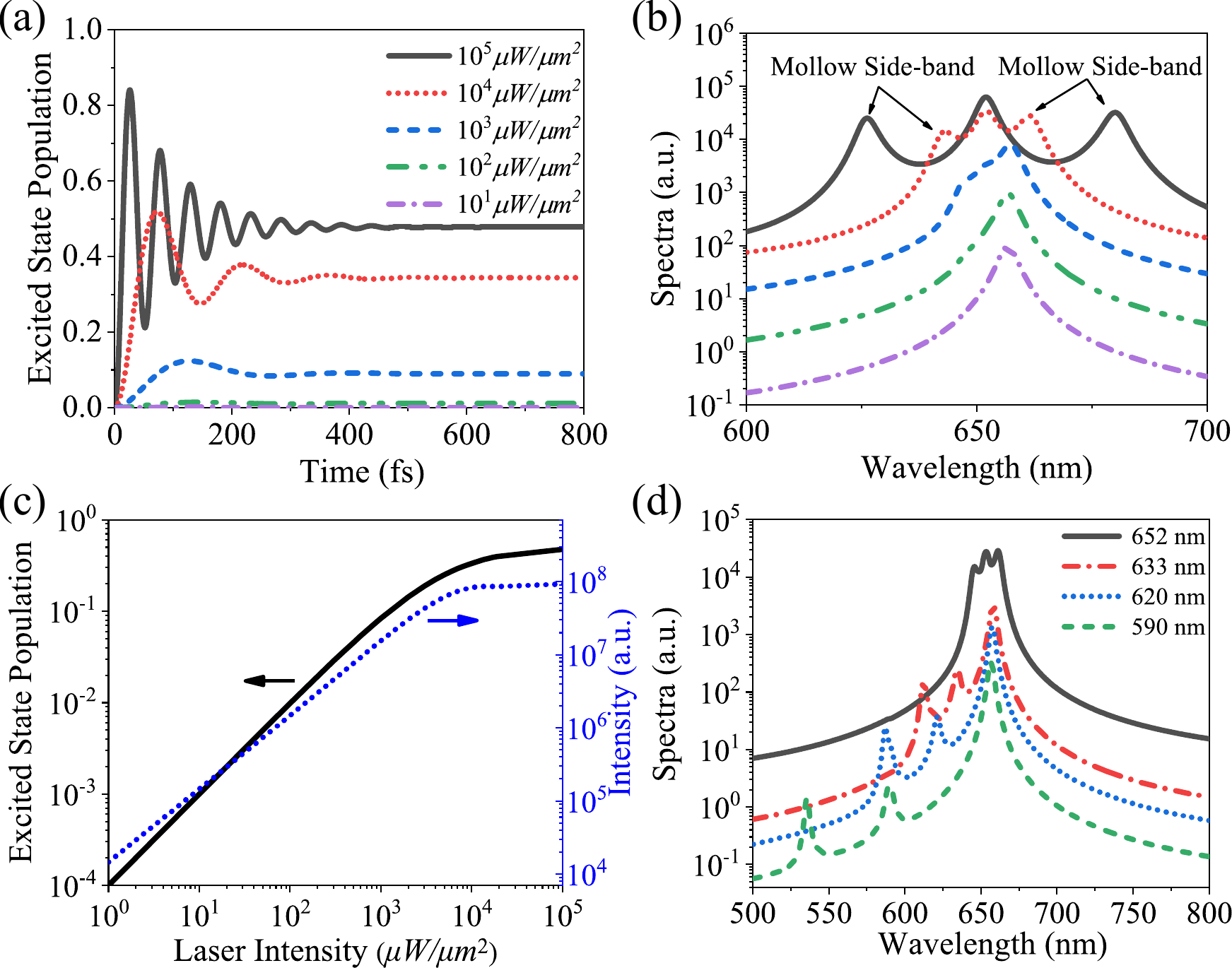}
\par\end{centering}
\caption{ \label{fig:6} Molecular dynamics and fluorescence. (a) and (b) show the dynamics of the excited state population and the fluorescence spectra of single ZnPc molecule, respectively, for the resonant CW laser excitation with wavelength $\lambda_{las} = 652$ nm and increasing intensity $I_{las}$  from $1 \mu W/\mu m^2$ to $10^5 \mu W/\mu m^2$.  (c) shows the steady-state population of the excited state (black solid line, left axis), and the integrated fluorescence intensity (blue dashed line, right axis) as a function of $I_{las}$  for  $\lambda_{las} =652$ nm. (d) shows the fluorescence spectrum for the laser excitation with varying $\lambda_{las}$ from $590$ nm to $652$  nm (from low to top curve) and $I_{las} = 10^4 \mu W/\mu m^2$. }
\end{figure*}

\subsection{Fluorescence Spectrum for Different Laser Wavelength and Intensity  \label{sec:fsintensity}}

So far, most of studies on the picocavity have focused on the sub-nanometre resolution of the optical imaging afforded by the extreme field confinement \citep{JLee,CCarnegie,HHShin,BDoppagne,BYang,ZHe,DuanS2015}. However, less attention has been paid to the extremely large field enhancement, which accompanies also with the picocavity. Thus, in this subsection, we explore how the large enhancement can affect the molecular dynamics and the fluorescence spectrum, see Fig. \ref{fig:6}.

Fig. \ref{fig:6}a,b show the dynamics of the excited state population $P_{e}=(\left< \sigma^z \right>+1)/2$ and the fluorescence spectrum for the 652 nm laser illumination with increasing  intensity  $I_{las}$. For $I_{las} = 10^3 \mu W/ \mu m^2$, $P_e$ increases firstly monotonously and then saturates, and the fluorescence spectrum shows single peak. For larger $I_{las}$, $P_{e}$  shows oscillatory behavior before reaching the saturated values, and the fluorescence spectrum shows three peaks. With further increased $I_{las}$, the period of the oscillations reduces, and the saturated population increases. In the meanwhile, the spectral intensity increases, and the peaks with large and small wavelength blue-shift and red-shift, respectively. These results indicate that for large  $I_{las}$, the molecule can be coherently excited to the superposition of the electronic ground and excited state, forming the so-called dressed states and leading to  the Mollow triplet in the spectrum \citep{MOScully,GeRC}. In Fig. \ref{fig:6}c, we  summarize the evolution of the saturated $P_{e}$  (black solid line, left axis) and the integrated fluorescence intensity $I_{flu}$  (blue dotted line, right axis) with increasing laser intensity  $I_{las}$. We see that  both $P_{e}$  and $I_{flu}$  increase linearly for small $I_{las}$, and then sub-linearly for moderate $I_{las}$, and finally saturate for larger $I_{las}$. 

Furthermore, we investigate with Fig. \ref{fig:6}d the influence of the laser wavelength $\lambda_{las}$ on the fluorescence spectrum for given laser intensity $I_{\rm las} = 10^5 \mu W/\mu m^2$. As $\lambda_{las}$ approaches the molecular resonance at $652$ nm, the spectrum intensity increases, and the Mollow side peaks with smaller wavelength blue-shift and finally merge with the dominated Mollow peak with larger wavelength. In the end, we emphasize that the laser intensity up to $I_{\rm las} = 10^4 \mu W/\mu m^2$ can be achieved with CW laser, and the much stronger intensity can be realized with the pulsed laser \citep{LombardiN}.   

\section{Conclusions}

In conclusion, to address the plasmonic picocavity-controlled fluorescence of single molecule, as demonstrated in the recent experiment \citep{BYang}, we have proposed a semi-classical theory by combining the macroscopic quantum electrodynamics theory and the open quantum system theory.  Our simulations not only reproduce the experimental observation, but also predict that the molecule can be coherently excited, and the Mollow triplets can be observed in the fluorescence spectrum  for sufficient strong laser illumination, which however can be achieved with CW or pulsed laser. Thus, our study highlights the possibility of  manipulating coherently the molecular states and exploring the non-linear optical phenomena with the plasmonic picocavity.

\section*{Acknowledgement}
We acknowledge project Nr. 12004344 from the National Natural Science Foundation of China, joint project Nr. 21961132023 from the NSFC-DPG. The calculations with Matlab and Gaussian $16$ were performed with the supercomputer at the Henan Supercomputer Center.

\section*{Author contributions}
Yuan Zhang has devised the theory. Yuan Zhang and Siyuan Lyu have developed the numerical code. Yao Zhang has carried out the TDDFT calculation, and Siyuan Lyu has carried out all the other simulations. All the authors contribute to the writing of the manuscript.

\appendix

\section{Plasmonic Lamb Shift and Purcell-enhanced Decay Rate  of Single Molecule in Plasmonic Picocavity \label{sec:molpic}}

According to the macroscopic quantum electrodynamics theory \citep{
NRivera,Scheel}, the electromagnetic field can be described as a continuum via the
Hamiltonian $\hat{H}_{f}=\int d^3\mathbf{r}\int_{0}^{\infty}d\omega_{f}\hbar\omega_{f}\hat{\mathbf{f}}^{\dagger}\left(\mathbf{r},\omega_{f}\right)\cdot\hat{\mathbf{f}}\left(\mathbf{r},\omega_{f}\right)
$ with frequency $\omega_{f}$, creation $\hat{\mathbf{f}}^{\dagger}\left(\mathbf{r},\omega_{f}\right)$ and annihilation $\hat{\mathbf{f}}\left(\mathbf{r},\omega_{f}\right)$ noise (bosonic) operators at position $\mathbf{r}$, and the quantized electric field operator is given by 
\begin{align}
 & \hat{\mathbf{E}}\left(\mathbf{r},\omega_{f}\right)=i\sqrt{\frac{\hbar}{\pi\epsilon_{0}}}\frac{\omega_{f}^{2}}{c^{2}}\int d^3\mathbf{r}'\nonumber \\
 & \times\sqrt{\epsilon^{I}\left(\mathbf{r}',\omega_{f}\right)}\overleftrightarrow{G}\left(\mathbf{r},\mathbf{r}';\omega_{f}\right)\cdot\hat{\mathbf{f}}\left(\mathbf{r}',\omega_{f}\right),\label{eq:Eoperator}
\end{align}
with the imaginary part of the dielectric function $\epsilon^{I}\left(\mathbf{r}',\omega\right)$
and the classical dyadic Green's function $\overleftrightarrow{G}\left(\mathbf{r},\mathbf{r}';\omega\right)$.
To study the interaction of single molecule with the picocavity, we model the molecule as two-level system via the Hamiltonian $\hat{H}_{m}=\left(\hbar\omega_{eg}/2\right)\hat{\sigma}^{z}$ with the transition frequency $\omega_{eg}$ and the Pauli operator $\hat{\sigma}^{z}$. In the rotating wave approximation, the molecule interacts with the quantized field via the Hamiltonian 
\begin{align}
 & \hat{H}_{fm}=-\int_{0}^{\infty}d\omega_{f}\Bigl[\hat{\sigma}^{\dagger}\int d^3\mathbf{r}\left(\frac{ie}{\omega_{eg}}\mathbf{j}_{eg}\left(\mathbf{r}\right)\right)\cdot\hat{\mathbf{E}}\left(\mathbf{r},\omega_{f}\right)\nonumber \\
 & +\int d^3\mathbf{r}\left(-\frac{ie}{\omega_{eg}}\mathbf{j}_{eg}^{*}\left(\mathbf{r}\right)\right)\cdot\hat{\mathbf{E}}^{\dagger}\left(\mathbf{r},\omega_{f}\right)\hat{\sigma}^{-}\Bigr].\label{eq:Interaction}
\end{align}

To proceed, we consider the electromagnetic field as reservoir, and apply the open quantum system theory to achieve an effective master equation for the molecule by adiabatically
eliminating the reservoir. To this end, we consider the Heisenberg
equation for the operator $\hat{o}$ of the molecule 
\begin{align}
 & \frac{\partial}{\partial t}\hat{o}\left(t\right)=\left[\hat{\sigma}^{\dagger}\left(t\right),\hat{o}\left(t\right)\right]\frac{ie}{\omega_{eg}c^{2}}\sqrt{\frac{1}{\hbar\pi\epsilon_{0}}}\int_{0}^{\infty}d\omega_{f}\omega_{f}^{2}\int d^3\mathbf{r}\nonumber \\
 & \times\int d^3\mathbf{r}'\sqrt{\epsilon^{I}\left(\mathbf{r}',\omega_{f}\right)}\mathbf{j}_{eg}\left(\mathbf{r}\right)\cdot\overleftrightarrow{G}\left(\mathbf{r},\mathbf{r}';\omega_{f}\right)\cdot\hat{\mathbf{f}}\left(\mathbf{r}',\omega_{f},t\right)\nonumber \\
 & +\frac{ie}{\omega_{eg}c^{2}}\sqrt{\frac{1}{\hbar\pi\epsilon_{0}}}\int_{0}^{\infty}d\omega_{f}\omega_{f}^{2}\int d^3\mathbf{r}\int d^3\mathbf{r}'\sqrt{\epsilon^{I}\left(\mathbf{r}',\omega_{f}\right)}\nonumber \\
 & \times\mathbf{j}_{eg}^{*}\left(\mathbf{r}\right)\cdot\overleftrightarrow{G}^{*}\left(\mathbf{r},\mathbf{r}';\omega_{f}\right)\hat{\mathbf{f}}^{\dagger}\left(\mathbf{r}',\omega_{f},t\right)\left[\hat{\sigma}^{-}\left(t\right),\hat{o}\left(t\right)\right].\label{eq:eq-o}
\end{align}
This equation depends on the field operator $\hat{\mathbf{f}}\left(\mathbf{r}',\omega_{f},t\right)$
(and its conjugation $\hat{\mathbf{f}}^{\dagger}\left(\mathbf{r}',\omega_{f},t\right)$), which follows the Heisenberg equation 
\begin{align}
 & \frac{\partial}{\partial t}\hat{\mathbf{f}}\left(\mathbf{r}',\omega_{f},t\right)=-i\omega_{f}\hat{\mathbf{f}}\left(\mathbf{r}',\omega_{f},t\right)-\frac{ie\omega_{f}^{2}}{\omega_{eg}c^{2}}\sqrt{\frac{\epsilon^{I}\left(\mathbf{r}',\omega_{f}\right)}{\hbar\pi\epsilon_{0}}}\nonumber \\
 & \times\int d^3\mathbf{r}\mathbf{j}_{eg}^{*}\left(\mathbf{r}\right)\cdot\overleftrightarrow{G}^{*}\left(\mathbf{r},\mathbf{r}';\omega_{f}\right)\hat{\sigma}^{-}\left(t\right),\label{eq:eqn-f}
\end{align}
where we have used the commutation relations  $\left[\hat{\mathbf{f}}\left(\mathbf{r}',\omega_{f}',t\right),\hat{\mathbf{f}}\left(\mathbf{r},\omega_{f},t\right)\right]=0$
and $\left[\hat{\mathbf{f}}^{\dagger}\left(\mathbf{r}',\omega'_{f},t\right),\hat{\mathbf{f}}\left(\mathbf{r},\omega_{f},t\right)\right]=-\delta\left(\mathbf{r}-\mathbf{r}'\right)\delta\left(\omega_{f}-\omega_{f}'\right)$.
The formal solution of Eq. (\ref{eq:eqn-f}) is 

\begin{align}
 & \hat{\mathbf{f}}\left(\mathbf{r}',\omega_{f},t\right)=-\frac{ie\omega_{f}^{2}}{\omega_{eg}c^{2}}\sqrt{\frac{\epsilon^{I}\left(\mathbf{r}',\omega_{f}\right)}{\hbar\pi\epsilon_{0}}} \int d^3\mathbf{r} \mathbf{j}_{eg}^{*}\left(\mathbf{r}\right)\nonumber \\
 & \cdot\overleftrightarrow{G}^{*}\left(\mathbf{r},\mathbf{r}';\omega_{f}\right)\int_{0}^{t}dt'e^{-i\omega_{f}\left(t-t'\right)}\hat{\sigma}^{-}\left(t'\right). \label{eq:feq-f}
\end{align}
The equation for the conjugate field operator $\hat{\mathbf{f}}^{\dagger}\left(\mathbf{r},\omega_{f},t\right)$
and its formal solution can be achieved by taking the conjugation
over Eq. (\ref{eq:eqn-f}) and (\ref{eq:feq-f}). 

Inserting Eq. (\ref{eq:feq-f}) into Eq. (\ref{eq:eq-o}), we will
obtain a differential and integral equation. By solving this equation,
we are able to study not only Markov dynamics in the weak coupling
regime, but also the non-Markov dynamics in the strong coupling regime \citep{TNeuman}.
Since here we focus on the former regime, we carry out the Born-Markov
approximation to the formal solution (\ref{eq:feq-f}). To do so,
we replace $\hat{\sigma}^{-}\left(\tau\right)$ by $e^{i\omega_{eg}\left(t-\tau\right)}\hat{\sigma}^{-}\left(t\right)$
in this expression, and then define a new variable $\tau=t-t'$ to
change the integration over the time, and finally change the upper
limit of this integration into infinity to achieve the following expression
\begin{align}
 & \hat{\mathbf{f}}\left(\mathbf{r}',\omega_{f},t\right)\approx-\frac{ie}{\omega_{eg}c^{2}\omega_{f}^{2}}\sqrt{\frac{\epsilon^{I}\left(\mathbf{r}',\omega_{f}\right)}{\hbar\pi\epsilon_{0}}}\int d^3\mathbf{r}\mathbf{j}_{eg}^{*}\left(\mathbf{r}\right)\nonumber \\
 & \cdot\overleftrightarrow{G}^{*}\left(\mathbf{r},\mathbf{r}';\omega_{f}\right)\hat{\sigma}^{-}\left(t\right)\left(\pi\delta\left(\omega_{eg}-\omega_{f}\right)+i\mathcal{P}\frac{1}{\omega_{eg}-\omega_{f}}\right).\label{eq:fBM}
\end{align}
In the last step, we have utilized the relationship 
$\int_{0}^{\infty}d\tau e^{i\left(\omega_{eg}-\omega_{f}\right)\tau}=\pi\delta\left(\omega_{eg}-\omega_{f}\right)+i\mathcal{P}\frac{1}{\omega_{eg}-\omega_{f}}$.
Inserting Eq. (\ref{eq:fBM}) (and its conjugation) into Eq. (\ref{eq:eq-o}),
using the property of the dyadic Green's function
\begin{align}
 & \left(\frac{\omega_{f}}{c}\right)^{2}\sum_{j}\int d^{3}\mathbf{r}'\epsilon^{I}\left(\mathbf{r}',\omega_{f}\right)G_{k'j}\left(\mathbf{r}_{1},\mathbf{r}';\omega_{f}\right)\nonumber \\
 & \times G_{kj}^{*}\left(\mathbf{r}_{2},\mathbf{r}';\omega_{f}\right)=\mathrm{Im}G_{k'k}\left(\mathbf{r}_{1},\mathbf{r}_{2};\omega_{f}\right),\label{eq:identity}
\end{align}
and applying the Kramer-Kronig relation 
\begin{align}
 & \mathcal{P}\int\frac{d\omega_{f}\omega_{f}^{2}}{\omega_{f}-\omega_{eg}}\mathrm{Im}\overleftrightarrow{G}\left(\mathbf{r},\mathbf{r}';\omega_{f}\right)=\pi\omega_{eg}^{2}\mathrm{Re}\overleftrightarrow{G}\left(\mathbf{r},\mathbf{r}';\omega_{eg}\right),\label{eq:KKrelation}
\end{align}
we obtain the following effective Heisenberg equation 
\begin{align}
\frac{\partial}{\partial t}\hat{o}\left(t\right) & =-i\left[\hat{\sigma}^{\dagger}\left(t\right),\hat{o}\left(t\right)\right]\hat{\sigma}^{-}\left(t\right)J^{\left(1\right)}\nonumber \\
 & -iJ^{\left(2\right)}\hat{\sigma}^{\dagger}\left(\tau\right)\left[\hat{\sigma}^{-}\left(t\right),\hat{o}\left(t\right)\right],\label{eq:emq}
\end{align}
with the spectral densities 
\begin{align}
J^{\left(1\right)} & =\frac{e^{2}}{\hbar\epsilon_{0}c^{2}}\int d^3\mathbf{r}\int d^3\mathbf{r}'\mathbf{j}_{eg}\left(\mathbf{r}\right)\cdot\overleftrightarrow{G}\left(\mathbf{r},\mathbf{r}';\omega_{eg}\right)\cdot\mathbf{j}_{eg}^{*}\left(\mathbf{r}'\right),\\
J^{\left(2\right)} & =\frac{e^{2}}{\hbar\epsilon_{0}c^{2}}\int d^3\mathbf{r}\int d^3\mathbf{r}'\mathbf{j}_{eg}\left(\mathbf{r}\right)\cdot\overleftrightarrow{G}^{*}\left(\mathbf{r},\mathbf{r}';\omega_{eg}\right)\cdot\mathbf{j}_{eg}^{*}\left(\mathbf{r}'\right).
\end{align}

Furthermore, we consider the equation for the expectation value
$\mathrm{tr}\left\{ \hat{o}\left(t\right) \hat{\rho}\right\} =\mathrm{tr}\left\{ \hat{o} \hat{\rho}\left(t\right)\right\} $,
which can be computed either with the time-dependent operator $\hat{o}\left(t\right)$
and the time-independent density operator $\hat{\rho}$ in the Heisenberg
picture (left side), or with the time-independent operator $\hat{o}$ and the
time-dependent density operator $\hat{\rho}\left(t\right)$ in the
Schr{\"o}dinger picture (right side). Using this relation and the cyclic property of
the trace, we obtain the master equation for the reduced density operator
\begin{align}
\frac{\partial}{\partial t}\hat{\rho} & =-iJ^{\left(1\right)}\left[\hat{\sigma}^{-}\hat{\rho},\hat{\sigma}^{\dagger}\right]-iJ^{\left(2\right)}\left[\hat{\rho}\hat{\sigma}^{\dagger},\hat{\sigma}^{-}\right]. \label{eq:meq-app}
\end{align}
Introducing new parameters $\Omega=J^{\left(1\right)}+J^{\left(2\right)},\Gamma=-i\left[J^{\left(1\right)}-J^{\left(2\right)}\right]$,
we can rewrite the spectral densities as $J^{\left(1\right)}=\frac{1}{2}\left(\Omega+i\Gamma\right),J^{\left(2\right)}=\frac{1}{2}\left(\Omega-i\Gamma\right).$
Inserting these expressions into Eq. (\ref{eq:meq-app}), we achieve the
following effective master equation 
\begin{align}
\frac{\partial}{\partial t}\hat{\rho} & =-i\frac{1}{2}\omega_{eg}\left[\hat{\sigma}^{z},\hat{\rho}\right]+i\frac{1}{2}\Omega\left[\hat{\sigma}^{\dagger}\hat{\sigma}^{-},\hat{\rho}\right]\nonumber \\
 & +\frac{1}{2}\Gamma\left(2\hat{\sigma}^{-}\hat{\rho}\hat{\sigma}^{\dagger}-\hat{\sigma}^{\dagger}\hat{\sigma}^{-}\hat{\rho}-\hat{\rho}\hat{\sigma}^{\dagger}\hat{\sigma}^{-}\right).\label{eq:master-equation}
\end{align}
Here, we have incorporated the molecular Hamiltonian $\hat{H}_{m}=\left(\hbar\omega_{eg}/2\right)\hat{\sigma}^{z}$. It is clear that $\Omega/2$ shifts the molecular transition, i.e. the plasmonic Lamb shift, and $\Gamma$ describes the excited state decay, i.e. the Purcell-enhanced decay rate. On the basis of this equation, we can further introduce the molecular coupling with the plasmon-enhanced local field, and the excited decay due to other processes and the molecular dephasing to arrive at the master equation (\ref{eq:meq}) in the main text.

\section{Far-field Spectrum }
In this Appendix, we present the derivation of the far-field radiation
from the single molecule in the picocavity. According to Refs. \citep{MOScully,DASteck}, the far-field spectrum can be computed with 
\begin{equation}
\frac{dW}{d\Omega}\left(\omega\right)=\frac{c\epsilon_{0}r^{2}}{4\pi^{2}}\mathrm{Re}\int_{0}^{\infty}e^{i\omega\tau}d\tau\mathrm{tr}\left\{ \hat{\mathbf{E}}^{\dagger}\left(\mathbf{r}_{d},0\right)\cdot\hat{\mathbf{E}}\left(\mathbf{r}_{d},\tau\right)\hat{\rho}\right\} .\label{eq:spectrum}
\end{equation}
In this expression, $r$ is the distance between the molecule and
the detector, $\hat{\mathbf{E}}\left(\mathbf{r}_{d},\tau\right)=\int d\omega_{f}\hat{\mathbf{E}}\left(\mathbf{r}_{d},\omega_{f},\tau\right)$
is the electric field operator at the detector position $\mathbf{r}_{d}$,
and is obtained by the integration of the electric field over the
frequency $\omega_{f}$. Inserting Eq. (\ref{eq:fBM}) into Eq. (\ref{eq:Eoperator}),
we obtain the following expression
\begin{align}
 & \hat{\mathbf{E}}\left(\mathbf{r}_{d},\omega_{f},\tau\right)\approx\frac{\hbar e}{\pi\omega_{eg}\epsilon_{0}}\frac{\omega_{f}^{2}}{c^{2}}\mathrm{Im}\int d^3\mathbf{r}'\overleftrightarrow{G}\left(\mathbf{r}_{d},\mathbf{r}';\omega_{f}\right)\nonumber \\
 & \cdot\mathbf{j}_{eg}^{*}\left(\mathbf{r}'\right)\hat{\sigma}^{-}\left(\tau\right)\left(\pi\delta\left(\omega_{f}-\omega_{eg}\right)+i\mathcal{P}\frac{1}{\omega_{eg}-\omega_{f}}\right).
\end{align}
Here, we have utilized the relation (\ref{eq:identity}). Using
the above expression and Eq. (\ref{eq:KKrelation}), we obtain the
following expression for the electric field operator at the position
$\mathbf{r}_{d}$ of the detector 

\begin{equation}
\hat{\mathbf{E}}\left(\mathbf{r}_{d},\tau\right)=-\frac{ie\omega_{eg}}{\epsilon_{0}c^{2}}\int d^3\mathbf{r}'\overleftrightarrow{G}\left(\mathbf{r}_{d},\mathbf{r}';\omega_{eg}\right)\cdot\mathbf{j}_{eg}^{*}\left(\mathbf{r}'\right)\hat{\sigma}^{-}\left(\tau\right).
\end{equation}

Applying the conjugation to the above equation, we obtain
the expression for the conjugated field operators $\hat{\mathbf{E}}^{\dagger}\left(\mathbf{r},\tau\right)$.
Inserting these results to Eq. (\ref{eq:spectrum}), we can rewrite
the spectrum as 
\begin{equation}
\frac{dW}{d\Omega}\left(\omega\right)\approx K\mathrm{Re}\int_{0}^{\infty}d\tau e^{i\omega\tau}\mathrm{tr}\left\{ \hat{\sigma}^{\dagger}\left(0\right)\hat{\sigma}^{-}\left(\tau\right)\hat{\rho}\right\}, \label{eq:spectrum-bm}
\end{equation}
with the propagation factors 
\begin{align}
K & =\frac{r^{2}\omega_{eg}^{2}e^{2}}{4\pi^{2}\epsilon_{0}c^{3}}\int d^3\mathbf{r}''\int d^3\mathbf{r}'\left[\overleftrightarrow{G}^{*}\left(\mathbf{r}_{d},\mathbf{r}'';\omega_{eg}\right)\cdot\mathbf{j}_{eg}\left(\mathbf{r}''\right)\right]\nonumber \\
 & \cdot\left[\overleftrightarrow{G}\left(\mathbf{r}_{d},\mathbf{r}';\omega_{eg}\right)\cdot\mathbf{j}_{eg}^{*}\left(\mathbf{r}'\right)\right].
\end{align}
To compute the spectrum with Eq. (\ref{eq:spectrum-bm}), we have
to evaluate the two-time correlations $\mathrm{tr}\left\{ \hat{\sigma}^{\dagger}\left(0\right)\hat{\sigma}^{-}\left(\tau\right)\hat{\rho}_{ss}\right\} $,
where $\tau$ labels the difference of time relative to the steady-state
labeled as ``$0$'', and $\hat{\rho}_{ss}$ is the density operator at steady-state. To compute these correlations, we consider
a pure quantum system. In this case, we can introduce the time-propagation
operator $\hat{U}\left(\tau\right)$ to reformulate the correlations
as 
\begin{align}
 & \mathrm{tr}\left\{ \hat{\sigma}^{\dagger}\left(0\right)\hat{\sigma}^{-}\left(\tau\right)\hat{\rho}_{ss}\right\} =\mathrm{tr}\left\{ \hat{\sigma}^{\dagger}\hat{U}^{\dagger}\left(\tau\right)\hat{\sigma}^{-}\hat{U}\left(\tau\right)\hat{\rho}_{ss}\right\} \nonumber \\
 & =\mathrm{tr}\left\{ \hat{\sigma}^{-}\hat{U}\left(\tau\right)\hat{\rho}\hat{\sigma}^{\dagger}\hat{U}^{\dagger}\left(\tau\right)\right\} =\mathrm{tr}\left\{ \hat{\sigma}^{-}\hat{\varrho}\left(\tau\right)\right\} ,
\end{align}
where we have defined the operator $\hat{\varrho}\left(\tau\right)=\hat{U}\left(\tau\right)\hat{\rho}_{ss}\hat{\sigma}^{\dagger}\hat{U}^{\dagger}\left(\tau\right)$.
In essence, we have transformed the expression in the Heisenberg picture
to that in the Schr{\"o}dinger picture. To deal with the quantum system in the presence of loss, we should replace $\hat{U}\left(\tau\right)...\hat{U}^{\dagger}\left(\tau\right)$
with the time-propagation superoperator $\hat{\mathcal{U}}\left(\tau\right)$,
which indicates the formal solution of the master equation (\ref{eq:meq}) in the main text with loss.
Finally, we can compute the spectrum as 
\begin{equation}
\frac{dW}{d\Omega}\left(\omega\right)\approx K\mathrm{Re}\int_{0}^{\infty}d\tau e^{i\omega\tau}\mathrm{tr}\left\{ \hat{\sigma}^{-}\hat{\varrho}\left(\tau\right)\right\} ,\label{eq:spec-Schordinger}
\end{equation}
where $\hat{\varrho}\left(\tau\right)$ satisfies the same master equation (\ref{eq:meq})
as $\hat{\rho}$ in the main text with however the initial condition $\hat{\varrho}\left(\tau\right)=\hat{\rho}_{ss}\hat{\sigma}^{\dagger}$.

\section{Supplemental Numerical Results \label{sec:snr}}
In this Appendix, we provide extra results to facilitate the discussions in the main text. 

\begin{figure}[!htb]
\begin{centering}
\includegraphics[scale=0.3]{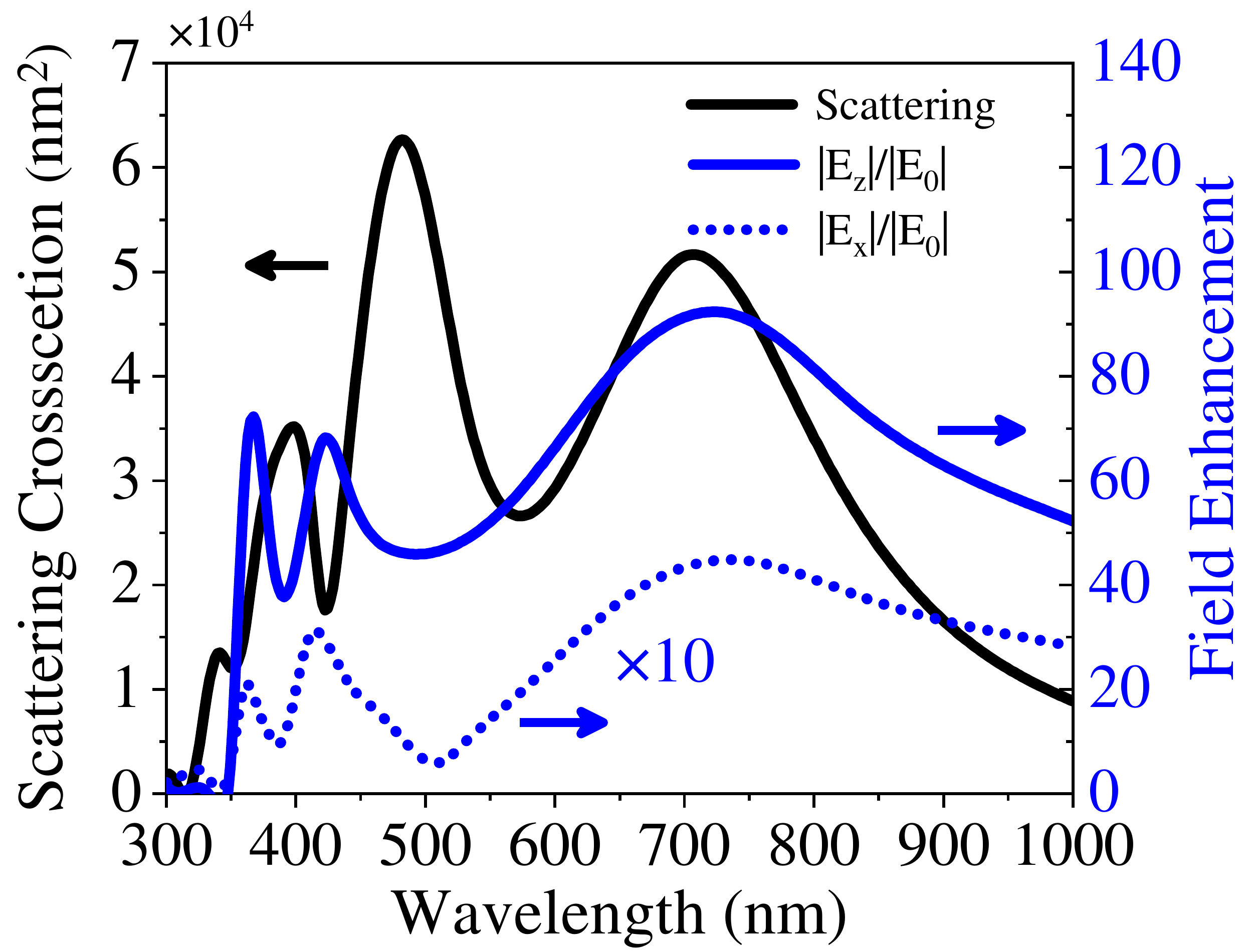}
\par\end{centering}
\caption{ \label{fig:2a_appendix} Plasmonic response of a nano-cavity: the far-field scattering cross-section (black solid line, left axis)  and the enhancement of the local field x- and z-component (blue dashed and solid line, right axis) as a function of wavelength of the plane-wave illumination. Here, the z- an x-component are evaluated at the molecular center and a point about $8.0$ nm away from the molecular center, respectively. }
\end{figure}

\subsection{Plasmonic Response of STM-based Nanocavity}

In the main text, we have studied the plasmonic response of the STM-based picocavity. As a comparison, here, we investigate the response of a STM-based nano-cavity, which resembles the picocavity except for the exclusion of the atomic protrusion. Fig. \ref{fig:2a_appendix} shows the computed far-field scattering cross section (black solid line, left axis), and the enhancement of the field z-component at the center of nanocavity (blue solid line), and of the x-component at a point $8.0$ nm away horizontally from the the center (blue dashed line). We find that the scattering spectrum is similar as that of the picocavity, indicating no influence of the atomic protrusion on the far-field field. In addition, the near field enhancement shows broad peaks with moderate values around $410$ nm in strong contrast to the sharp and strong peak at $430$ nm in the pico-cavity case (see Fig. \ref{fig:2}a).

\begin{figure}[!htb]
\begin{centering}
\includegraphics[scale=0.7]{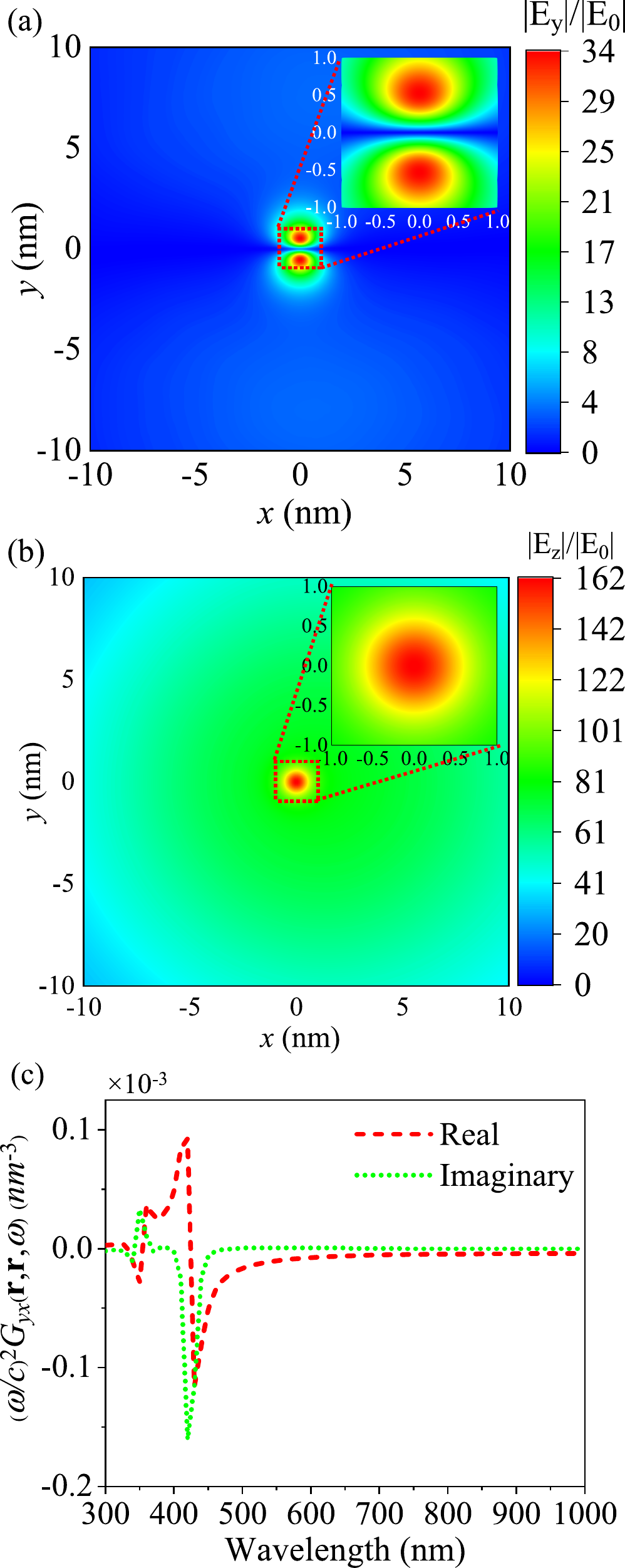}
\par\end{centering}
\caption{ \label{fig:2b_appendix} Local field mapping and dyadic Green's function. (a,b) show the  mapping of the local field y-component (a) and z-component (b) in the middle of the picocavity for a plane-wave illumination of $633$ nm wavelength. (c) shows the real part (red dashed line) and imaginary part (green dotted line)  of the scattered dyadic Green's function for the STM tip about $0.5$ nm away horizontally from the molecule center.}
\end{figure}

\subsection{Extra Near-field Mapping and Green's Function Component}

In the main text, we have characterized the main components of the near field and the dyadic Green's function for the plasmonic pico-cavity. Here, we provide the corresponding results for other components. Fig. \ref{fig:2b_appendix}a and b show the mapping of the local field y-component (a) and z-component (b) in the middle plane of the picocavity for the $633$ nm plane-wave illumination. The y-component field mapping is similar to the x-component field mapping except that the maximum occurs along the y-axis. The field z-component concentrates at the origin in an area of $1$ nm size over a broad background of $10$ nm size. Fig. \ref{fig:2b_appendix}c shows the real part (red dashed line) and the imaginary part (green dotted line) of the yx-component of the dyadic Green's tensor. The imaginary part shows a dip at around $420$ nm and a peak at around $350$ nm. There, we see also two Fano-features around $420$ nm and $360$ nm in the real part (red dashed line) of the dyadic Green's function.

\begin{figure}
\begin{centering}
\includegraphics[scale=0.8]{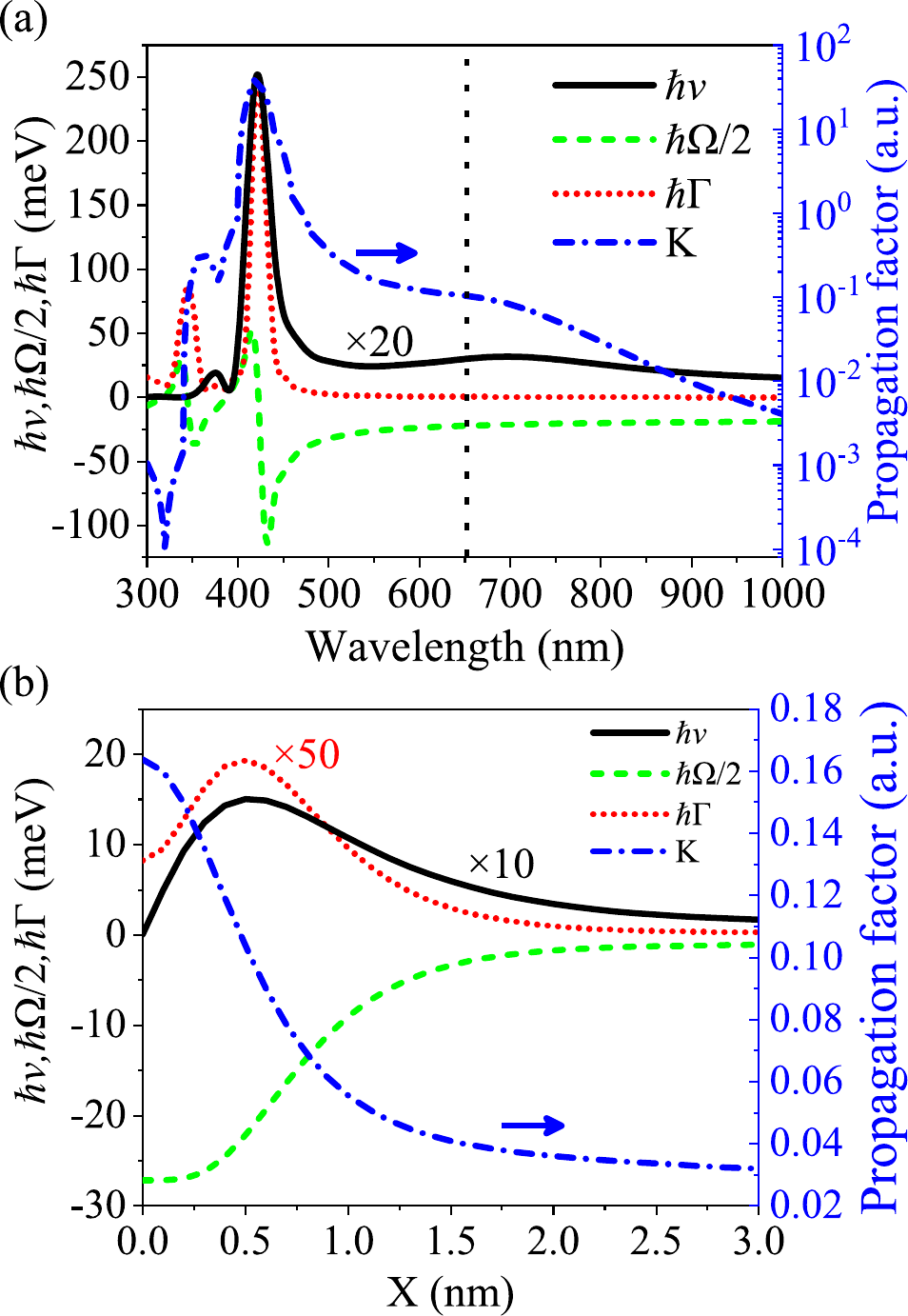}
\par\end{centering}
\caption{\label{fig:4_appendix} Response within dipole approximation. (a) shows the molecule-local field coupling $\hbar v$ (black solid line) for the input power $10^2 \mu W/\mu m^2$, the plasmonic Lamb shift $\hbar \Omega/2$  (blue dashed line), and the Purcell-enhanced decay rate $\hbar\Gamma$  (red dotted line), and the propagation factor $K$ (dash-dotted line) as a function of wavelength for the STM tip about $0.5$  nm away from the molecule center, where the vertical black dashed line shows the wavelength of the molecular transition. (b) shows the change of  $\hbar v$,  $\hbar \Omega/2$, $\hbar\Gamma$, $K$ as the STM tip moves away from the molecular center along the x-axis for the wavelength of molecular transition.   }
\end{figure}

\subsection{Molecule-Picocavity Coupling and Picocavity-controlled Fluorescence within Dipole Approximation}

In the main text, we have studied extensively the coupling and the fluorescence with the model accounting for the atomistic detail of the molecule. As a comparison, here, we study the results within the dipole approximation, which in principle is not valid for the system considered here.  

In Fig. \ref{fig:4_appendix}a, we calculate the molecule-local field coupling $\hbar v$ for given laser intensity  $I_{\rm las} = 10^2 \mu W/\mu m^2$ (black solid line), the plasmonic Lamb shift $\hbar \Omega/2$ (blue dashed line) and the Purcell-enhanced decay rate $\hbar\Gamma$ (red dotted line), and the propagation factor $K$ (green dash-dotted line) by modelling the molecule as a point within the dipole approximation. $\hbar v$ follows the shape of the near-field enhancement, and reaches the maximal value around $12.5$ meV at the wavelength of $420$ nm. $\hbar \Omega/2$ follows the shape of the real part of the dyadic Green's function, but changes in the range of $[-120,0  {\rm meV}]$. $\hbar \Gamma$ follows the shape of the imaginary part of that function, and varies in the range of $[0,250 {\rm meV}]$.

In Fig. \ref{fig:4_appendix}b, we investigate how the molecule-local field coupling, the plasmonic Lamb shift and the Purcell-enhanced decay rate change with the STM tip-molecule horizontal distance X. We see that the absolute value of these quantities increases first and then decreases with the increase of the distance X. More precisely, the molecule-local field coupling, the plasmonic Lamb shift and Purcell-enhanced decay rate reach their maximum $1.5$ meV, $-26$ meV, $ 0.4$ meV for the distance X around $0.5,0,0.5$ nm, respectively. The propagation factor reaches the maximum at the distance X $=0$ nm.

\subsection{Picocavity-controlled Fluorescence of the ${\rm S}_0 \to {\rm S}_1$ Transition}

\begin{figure}
\begin{centering}
\includegraphics[scale=0.28]{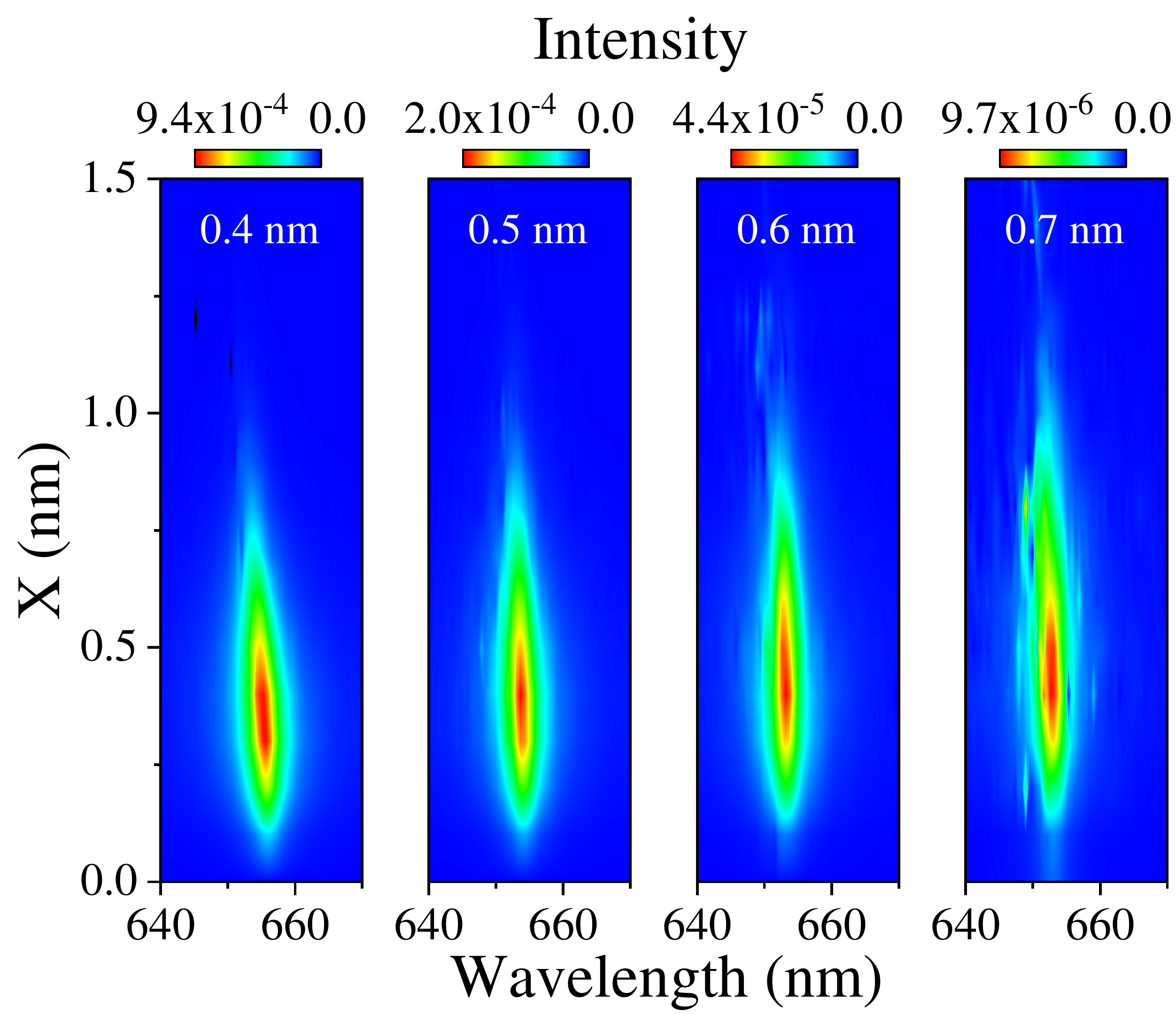}
\par\end{centering}
\caption{\label{fig:5_appendix} Picocavity-controlled fluorescence of single ZnPc molecule for ${\rm S}_0$ $\to$ ${\rm S}_1$ transition. Here, we utilize dephasing rate $\hbar \chi = 2.3$ meV, the intrinsic decay rate $ \hbar \gamma = 8.2$ meV.}
\end{figure}

Fig. \ref{fig:5_appendix} shows the evolution of fluorescence spectra of the ${\rm S}_0 \to {\rm S}_1$ transition as a function of the STM tip-molecule horizontal distance X for the increasing vertical distance $d_{tip}$  from $0.4$ nm to $0.7$ nm (from left to right panels). We find that it is five orders of magnitude smaller than the intensity in Fig. \ref{fig:5}a, because the propagation factor is orders of magnitude smaller for this transition under the detection scheme as considered here. Note that the two transitions are both excited to the same degree when the STM tip is in the middle of the molecule due to the field x- and y-component with similar strength, while the ${\rm S}_0$ $\to$ ${\rm S}_2$ transition is preferentially excited when the STM tip moves along the x-axis. Note that the noise feature in the spectra of the right two columns is due to the numerical issue during the Fourier transformation.

\end{document}